\documentclass[10pt]{article}
\newcommand{\llangle}{\langle\!\langle}
\newcommand{\rrangle}{\rangle\!\rangle}
\newcommand{\LLangle}{\left\langle\!\!\!\left\langle}
\newcommand{\RRangle}{\right\rangle\!\!\!\right\rangle}
\begin{document}

\makeatletter
\@addtoreset{equation}{section}
\makeatother
\renewcommand{\theequation}{\thesection.\arabic{equation}}
\baselineskip 15pt %<---- Toglierlo se si vuole il double-space
%\footnotesep 7mm % <------ Aumenta separazione tra footnotes

\title{\bf Dynamical Reduction Models with General Gaussian Noises}

\author{Angelo Bassi\footnote{e-mail: bassi@ictp.trieste.it}\\
{\small The Abdus Salam International Centre for
Theoretical Physics, Trieste, Italy,}\\ and \\
\\ GianCarlo Ghirardi\footnote{e-mail: ghirardi@ts.infn.it}\\
{\small Department of Theoretical Physics of the University of
Trieste, and}\\ {\small the Abdus Salam International Centre for
Theoretical Physics, Trieste, Italy.}}

\date{}

\maketitle

\begin{abstract}

We consider the effect of replacing in stochastic differential
equations
leading to the dynamical collapse of the statevector, white noise
stochastic processes with non white ones. We prove that such a
modification can be consistently performed without altering the most
interesting features of the previous models. One of the reasons to
discuss this matter derives from the desire of being allowed to deal
with
physical stochastic fields, such as the gravitational one, which
cannot
give rise to white noises. From our point of view the most relevant
motivation for the approach we propose here derives from the fact that
in relativistic models the occurrence of white noises is the main
responsible for the appearance of untractable divergences. Therefore,
one can hope that resorting to non white noises one can overcome such
a
difficulty. We investigate stochastic equations with non white
noises, we discuss their reduction properties and their physical
implications. Our analysis has a precise interest not only for the
above
mentioned subject but also for the general study of dissipative
systems
and decoherence.
\end{abstract}

\section{Introduction}

The aim of dynamical reduction models \cite{grw}---\cite{gis1} is
to combine the {\it Schr\"o\-din\-ger evolution} and the {\it
wavepacket reduction postulate} into one universal dynamical
equation, which is assumed to govern all physical processes. In
this way, such a dynamics accounts both for the quantum properties
of microscopic systems and for the classical properties of
macroscopic ones.

This goal is achieved by adding to the Schr\"odinger equation new
stochastic terms which induce a diffusion process --- guided by a
set of Gaussian white noises --- of the statevector in Hilbert
space: it is precisely this sort of ``random walk'' which is
responsible for the localization mechanism. We will review these
attempts in section 2.

The main aim of this paper is to generalize the basic equations of
dynamical reduction models to the case of general Gaussian noises,
not necessarily white in time. There are three reasons for
considering models of this kind: first of all, it is interesting
to analyze if, and to which extent, the nice features of the
reduction mechanisms proposed up to now depend on the white--noise
character of the stochastic processes. Secondly, it has been
argued several times that the stochastic processes should be
related to {\it physical} fields --- the most promising being the
gravitational field; in such a case, the stochastic processes
cannot be white, since white noises are never realized in nature.
Finally, the third motivation to look for generalizations of the
previous models has to do with the important problem of working
out relativistic models of wavepacket reduction \cite{ggp}. In
fact, even though it has been proved that models of this type
share all the nice features of the non relativistic ones, they
suffer from the serious drawback of inducing an infinite increase
of the energy of physical systems. This infinite increase is
basically caused by the local coupling between the quantum fields
and the white--noise stochastic fields appearing in such theories.
Accordingly, a reasonable way to overcome such a difficulty is to
replace the white--noise fields with more general ones: this will
be the subject of papers in preparation. Such attempts, however,
require a preliminary and detailed investigation aimed to clarify
that the consideration of noises which are non white in time does
not lead to inconsistencies and preserves the nice features of the
models based on noises which are white. The analysis we are going to
perform is also of interest, {\it per se}, for the study of stochastic
dynamical equations in general, a subjet which has received a lot of
attention in recent times.

Sections 3 to 6 of this paper are devoted to a general discussion
of non--white random differential equations in Hilbert space. We
will show that the most important features of white--noise
stochastic equations, in particular the desired reduction effects,
hold also in the more general case, thus proving that the
white--noise character of the stochastic processes is not an
essential element of the dynamical reduction program. Dynamical
reduction models based on general Gaussian noises have been
studied  by Pearle \cite{p1, p2} and Di\'osi {\it et al} \cite{g4,
gi2} along different lines. In the final section of the paper, we
apply the results and the formalism of the previous sections to a
specific model of dynamical reduction and we prove that it leads
precisely to the localization of macroscopic objects in space.

\section{Review of CSL}

The CSL (Continuous Spontaneous Localizations) version of
Dynamical Reduction Models \cite{pp1, gpr} is based on a
stochastic Schr\"odinger equation which, in the Stratonovich
language, takes the form:
\begin{equation} \label{csl1}
\frac{d|\psi(t)\rangle}{dt} \quad = \quad \left[ -\frac{i}{\hbar}
H_{0}\; +\; \sum_{i} A_{i}w_{i}(t)\; -\; \gamma \sum_{i}
A^{2}_{i}\right] |\psi(t)\rangle.
\end{equation}
Here, $H_{0}$ is the free Hamiltonian of the system; $\{ A_{i} \}$
is a set of commuting self--adjoint operators (representing the
preferred basis) whose common eigenmanifolds are the linear
manifolds into which the statevectors of individual physical
systems are driven; $w_{i}(t)$ are $c$--number independent
stochastic processes with a gaussian distribution which is white
in time:
\begin{equation} \label{wn}
\llangle w_{i}(t) \rrangle = 0, \qquad \quad
\llangle w_{i}(t_{1}) w_{j}(t_{2})\rrangle = \gamma\, \delta_{ij}\,
\delta(t_{1} - t_{2}),
\end{equation}
the symbol $\llangle . \rrangle$ denoting the stochastic
average associated to the process (\ref{wn}). Equation
(\ref{csl1}) describes a diffusion process in Hilbert space; it is
a linear equation like the Schr\"odinger equation, but it does not
preserve the norm of $|\psi(t)\rangle$ since the evolution is not
unitary, due to the presence of the last two terms on the right
hand side. The solution $|\psi(t)\rangle$ cannot therefore be
endowed with a direct physical meaning.

To overcome this difficulty, and at the same time to ensure that
the reduction mechanism reproduces the quantum mechanical
probabilities, the following strategy has been adopted \cite{pp1,
gpr}. The {\it physical vectors} are the normalized solutions of
equation (\ref{csl1}):
\begin{equation} \label{pv}
|\psi_{\makebox{\tiny Phys}}(t)\rangle \quad = \quad
\frac{|\psi(t)\rangle}{\| |\psi(t)\rangle \|},
\end{equation}
and it is assumed that any particular realization of the
stochastic processes $w_{i}(t)$, yielding the state
$|\psi_{\makebox{\tiny phys}}(t)\rangle$, has a probability of
occurrence $P_{\makebox{\tiny Cook}}[w(t)]$ equal to:
\begin{equation} \label{prob}
P_{\makebox{\tiny Cook}}[w(t)] \quad = \quad
P_{\makebox{\tiny Raw}}[w(t)]\,
\||\psi(t)\rangle\|^{2},
\end{equation}
where $P_{\makebox{\tiny Raw}}[w(t)]$ is the original probability
distribution of the gaussian white noises given by (\ref{wn}).

Of course, since (\ref{prob}) defines a probability distribution,
it must sum to 1:
\begin{eqnarray} \label{nor}
\int {\mathcal D}[w(t)]\; P_{\makebox{\tiny Cook}}[w(t)] & = & 1 \\
& = & \int {\mathcal D}[w(t)]\; P_{\makebox{\tiny Raw}}[w(t)]\,
\||\psi(t)\rangle\|^{2} \nonumber \\
& = & \llangle \langle\psi(t)|\psi(t)\rangle \rrangle
\nonumber
\end{eqnarray}
(we remember that $\llangle . \rrangle$ refers to the average
with respect to the original probability distribution
$P_{\makebox{\tiny Raw}}[w(t)]$). Equation (\ref{nor}) imposes
that the stochastic average of the square norm of the vector
$|\psi(t)\rangle$ be conserved; as it can be easily verified,
equation (\ref{csl1}) guarantees that this is the case.

It is possible to write \cite{gpr} the norm--preserving equation
for the physical vector $|\psi_{\makebox{\tiny Phys}}(t)\rangle$
which is equivalent to eq. (\ref{csl1}) with the prescription
(\ref{prob}): such an equation is nonlinear and more difficult to
handle. We will not consider it here.

In reference \cite{gpr} it has been shown that, if one ignores the
Hamiltonian term $H_{0}$, equation (\ref{csl1}) together with the
cooking prescription (\ref{prob}) --- or, alternatively, the
corresponding norm--preserving equation --- drives the statevector
of any individual physical system into one of the common
eigenmanifolds of the operators $A_{i}$. Here we will simply show
that equation (\ref{csl1}) implies the diagonalization of the
density matrix with respect to the basis of the common
eigenmanifolds of the operators\footnote{Of course, the
diagonalization of the density matrix is only a necessary, {\it
not} a sufficient condition for the localizations to occur, as it
has been shown in reference \cite{gnd}. Anyway, as already
remarked, reference \cite{gpr} contains a proof that equation
(\ref{csl1}) does imply the reduction of the statevector into the
desired eigenmanifolds.} $A_{i}$.

The statistical operator is the average value, with respect to the
the ``cooked'' (i.e. the physical) probability distribution
$P_{\makebox{\tiny Cook}}[w(t)]$, of the projections operators
onto the one dimensional linear manifolds spanned by the physical
vectors $|\psi_{\makebox{\tiny Phys}}(t)\rangle$ :
\begin{eqnarray} \label{rcdos}
\rho(t) & = & \int {\mathcal D}[w(t)]\,
|\psi_{\makebox{\tiny Phys}}(t)\rangle \langle
\psi_{\makebox{\tiny Phys}}(t)|\,
P_{\makebox{\tiny Cook}}[w(t)] \nonumber \\
& = &
\llangle |\psi(t)\rangle \langle\psi(t)| \rrangle.
\end{eqnarray}
We note that, from the {\it mathematical} point of view, $\rho(t)$
corresponds also to the ensemble of operators
$|\psi(t)\rangle\langle\psi(t)|$ ($|\psi(t)\rangle$ being non
normalized), averaged with the raw probability distribution
$P_{\makebox{\tiny Raw}}[w(t)]$. Thanks to this property the
dynamical evolution equation for $\rho(t)$ can be easily derived:
\begin{equation} \label{cslso}
\frac{d\rho(t)}{dt} \quad = \quad -\frac{i}{\hbar}
\left[ H_{0}, \rho(t) \right]\; - \;
\frac{\gamma}{2}\sum_{i} \left[A_{i}, \left[A_{i},\, \rho(t)
\right]\right].
\end{equation}

To show the effect of the reducing terms, let us suppose, for
simplicity, that the common eigenmanifolds $M_{\alpha}$ of the
operators $A_{i}$ (which we assume to have a purely discrete
spectrum) are one--dimensional and we call $|\alpha\rangle$ the
vector spanning $M_{\alpha}$:
\begin{equation}
A_{i}\, |\alpha\rangle \quad  = \quad a_{i\alpha}\,
|\alpha\rangle.
\end{equation}
For the moment, let us ignore the Hamiltonian term $H_{0}$.
Then, equation (\ref{cslso}) implies the following equation for the
matrix elements $\langle\alpha| \rho(t) |\beta\rangle$:
\begin{equation} \label{rm}
\frac{d\langle\alpha| \rho(t) |\beta\rangle}{dt} \quad = \quad
-\, \frac{\gamma}{2} \, \sum_{i} \left(a_{i\alpha}\, - \, a_{i\beta}
\right)^{2}\, \langle\alpha| \rho(t) |\beta\rangle.
\end{equation}
Equation (\ref{rm}) shows that the off--diagonal elements of the
density matrix, corresponding to the interference terms arising
from the superpositions of different eigenstates of $A_{i}$, are
exponentially damped. The diagonal elements, on the other hand, do
not change with time.

Up to now we have described the general formal structure of CSL. To
give a physical content to the model, one must choose the
``preferred basis'', i.e. the operators $A_{i}$ which define the
manifolds onto which the wavefunction is reduced. Obviously, our
aim is to induce the spatial localization of macroscopic objects.
To this purpose one can make the choice \cite{gpr}:
\begin{equation}\label{cho2}
A_{i} \quad\longrightarrow\quad {\mathcal N}({\bf x})\; = \;
\left(\frac{\alpha}{2\pi}\right)^{\frac{3}{2}} \sum_{s}
\int d^{\, 3}y\;
e^{\displaystyle -\frac{\alpha}{2}({\bf x} - {\bf y})^{2}}
a^{\dagger}({\bf y}, s)\,a({\bf y}, s),
\end{equation}
where $a^{\dagger}({\bf y}, s)$ and $a({\bf y}, s)$ are the
creation and annihilation  operators for a constituent with spin
component $s$, at point\footnote{If more than one type of particle
is involved, an extra sum over the different kinds of particles
must appear in the definition of ${\mathcal N}({\bf x})$.} ${\bf
y}$. The parameter $1/\sqrt{\alpha}$ is a measure the localization
accuracy of the reducing mechanism: for physical reasons
\cite{grw} its value has been chosen to be $1/\sqrt{\alpha} \simeq
10^{-5}$cm. The value of the other parameter of the theory,
$\gamma$, which measures the strength of the correlation function
of the white noises, is related to the parameter $\lambda \simeq
10^{-16}$ sec$^{-1}$ of QMSL \cite{grw}, which, within such a discrete
model, specifies the  frequency of the random reduction processes,
according to
$\gamma =
\lambda (4\pi/\alpha)^{3/2}$. Accordingly, the stochastic processes
$w_{i}(t)$ are replaced by a gaussian stochastic field $w({\bf x},
t)$, whose first two moments are:
\begin{equation} \label{wn2}
\llangle w({\bf x}, t) \rrangle = 0, \qquad \quad
\llangle w({\bf x}, t_{1}) w({\bf y}, t_{2})\rrangle = \gamma\,
\delta({\bf x} - {\bf y})\, \delta(t_{1} - t_{2}).
\end{equation}
The modified Schr\"odinger equation (\ref{csl1}) becomes then:
\begin{equation} \label{sch1b}
\frac{d|\psi(t)\rangle}{dt} \quad = \quad \left[ -\frac{i}{\hbar}
H_{0}\; +\; \int d^{\,3}x\; {\mathcal N}({\bf x}) w({\bf x},t)
\; - \; \gamma \int d^{\,3}x\; {\mathcal N}^{2}({\bf x})  \right]
|\psi(t)\rangle,
\end{equation}
and the corresponding equation for the statistical operator is:
\begin{equation} \label{cslso2}
\frac{d\rho(t)}{dt} \; = \; -\frac{i}{\hbar} \left[ H_{0}, \rho(t)
\right]
\; - \; \frac{\gamma}{2}\,  \int d^{\,3}x\;
\left[ {\mathcal N}({\bf x}), \left[
{\mathcal N}({\bf x}), \rho(t) \right]\right].
\end{equation}
Equations (\ref{sch1b}) and (\ref{cslso2}) exhibit two basic
features:
\begin{itemize}
  \item At the microscopic level --- i.e. when only few
constituents are involved --- the new terms do not alter in any
appreciable way the pure Schr\"odinger evolution: all quantum
properties of micro--systems are left essentially unchanged.
This is due to the fact that the value of $\lambda$ is
extremely small.
  \item At the macroscopic level, on the other hand, the new
terms induce in a very short time --- much shorter than the
perception time of a conscious observer --- the suppression of the
superposition of different macroscopic states and the reduction to
one of them. Accordingly, macroscopic objects are always localized
in space, and their classical properties are restored.
\end{itemize}
This is the way in which dynamical reduction models are able to
account for the behaviour of both microscopic quantum
and macroscopic classical systems.

\section{Dynamical reduction models with general Ga\-us\-sian
noises}

In this section we begin the analysis of dynamical reduction
models in which the reduction mechanism is controlled by general
Gaussian noises. The first task is to derive a modified
Schr\"odinger equation generalizing equation (\ref{csl1}), and
preserving the average value of the square norm of vectors, so
that the cooking prescription can be applied to it.

Let us then consider the following equation:
\begin{equation} \label{nm1}
\frac{d|\psi(t)\rangle}{dt} \quad = \quad \left[ -\frac{i}{\hbar}
H_{0}\; +\; \sum_{i} A_{i}w_{i}(t) \right] |\psi(t)\rangle,
\end{equation}
where, as before, $H_{0}$ is the Hamiltonian of the system, $\{
A_{i} \}$ is a set of commuting self--adjoint operators, and
$w_{i}(t)$ are $c$--number gaussian stochastic processes whose
first two moments are\footnote{There is no loss of generality in
considering gaussian processes with zero mean. In fact, if
$\llangle w_{i}(t) \rrangle = m_{i}(t) \neq 0$, we can always
define new processes $z_{i}(t) = w_{i}(t) - m_{i}(t)$, which have
zero mean, and rewrite the Schr\"odinger equation (\ref{nm1}) in
terms of the processes $z_{i}(t)$.}:
\begin{equation} \label{cn}
\llangle w_{i}(t) \rrangle = 0, \qquad \quad
\llangle w_{i}(t_{1}) w_{j}(t_{2})\rrangle =
\gamma\,D_{ij}(t_{1}, t_{2}).
\end{equation}
As in standard CSL, the evolution described by equation (\ref{nm1})
is not unitary and it does not preserve the norm of the
statevector; we then follow the same prescription outlined in
section 2. We consider as {\it physical} vectors the normalized
ones:
\begin{equation} \label{co2}
|\psi_{\makebox{\tiny Phys}}(t)\rangle \quad = \quad
\frac{|\psi(t)\rangle}{\| |\psi(t)\rangle \|},
\end{equation}
and we assume that any particular realization of the stochastic
processes $w_{i}(t)$  has a probability of occurrence
$P_{\makebox{\tiny Cook}}[w(t)]$ equal to:
\begin{equation} \label{prob2}
P_{\makebox{\tiny Cook}}[w(t)] \quad = \quad
P_{\makebox{\tiny Raw}}[w(t)]\, \||\psi(t)\rangle\|^{2},
\end{equation}
where $P_{\makebox{\tiny Raw}}[w(t)]$ is now the gaussian
probability distribution defined by (\ref{cn}).  The above
assumptions guarantee, as we will show in section 5, that the
reduction probabilities reproduce standard quantum mechanical
probabilities.

As in section 2, we have to impose that equation (\ref{prob2})
correctly defines a probability distribution, i.e. that it sums to
1. From equation (\ref{nor}) we see that this is equivalent to
requiring that the time derivative of $\llangle
\langle\psi(t)|\psi(t)\rangle \rrangle$ is zero. Let us evaluate
it:
\begin{eqnarray*}
\frac{d}{dt}\llangle\langle\psi(t)|\psi(t)\rangle\rrangle & = &
\LLangle\left[\frac{d\langle\psi(t)|}{dt}\right]|\psi(t)
\rangle\RRangle\; + \;
\LLangle\langle\psi(t)|\left[\frac{d|\psi(t)\rangle}{dt}\right]
\RRangle\quad = \\
& = & \LLangle\langle\psi(t)|\left[ +\;\frac{i}{\hbar}
H_{0}\; + \;\sum_{i}
A_{i}w_{i}(t)\right]|\psi(t)\rangle\RRangle \; + \\
& &  \LLangle\langle\psi(t)|\left[ -\;\frac{i}{\hbar}H_{0}
\; + \;\sum_{i}
A_{i}w_{i}(t)\right]|\psi(t)\rangle\RRangle.
\end{eqnarray*}
The two terms involving the Hamiltonian $H_{0}$ cancel out (in
fact they describe the unitary part of the evolution); the noises
$w_{i}(t)$, being $c$--numbers,
can be taken out of the scalar product, so that:
\begin{equation} \label{ie1}
\frac{d}{dt}\llangle\langle\psi(t)|\psi(t)\rangle\rrangle \quad =
\quad 2 \sum_{i} \llangle\langle\psi(t)|A_{i}|\psi(t)\rangle
w_{i}(t)\rrangle.
\end{equation}
The right hand side of (\ref{ie1}) can be rewritten with the help
of the {\it Furutsu--Novikov formula} \cite{ks}:
\begin{equation} \label{fn}
\llangle F[w(t)]w_{i}(t) \rrangle \quad = \quad
\gamma \sum_{j} \int_{0}^{+\infty}
D_{ij}(t,s) \LLangle \frac{\delta F[w(t)]}{\delta w_{j}(s)}
\RRangle\, ds
\end{equation}
(for simplicity, throughout this section we take $t_{0} = 0$ as the
initial time). $F[w(t)]$ is any functional of the stochastic fields
$w_{i}(t)$; in the present case case, $F[w(t)] =
\langle\psi(t)|A_{i}|\psi(t)\rangle$.

The formal solution of equation (\ref{nm1}) is:
\begin{equation} \label{fs}
|\psi(t)\rangle = T\, e^{\displaystyle
-\, \frac{i}{\hbar}H_{0} t\, + \sum_{i} A_{i}
\int_{0}^{t} w_{i}(s)\, ds }|\psi(0)\rangle.
\end{equation}
Note that, since $|\psi(t)\rangle$ depends on the stochastic
processes $w_{i}(s)$ only within the time--interval $[0,t]$, the
functional derivative of $|\psi(t)\rangle$ with respect to
$w_{j}(s)$ is zero if $s \not\in [0,t]$. We then have:
\begin{eqnarray} \label{ie2}
\frac{d}{dt}\llangle\langle\psi(t)|\psi(t)\rangle\rrangle & = &
2\gamma \sum_{i,j} \int_{0}^{t} D_{ij}(t,s)
\LLangle \left[ \frac{\delta \langle\psi(t)|}{\delta
w_{j}(s)}\right] A_{i}|\psi(t)\rangle\RRangle \,ds
\nonumber \\ &  &  \\ & + &
2\gamma \sum_{i,j} \int_{0}^{t} D_{ij}(t,s)
\LLangle \langle\psi(t)|A_{i} \left[ \frac{\delta
|\psi(t)\rangle}{\delta w_{j}(s)} \right]\RRangle \, ds
\; \neq \; 0. \nonumber
\end{eqnarray}
Since the time derivative of the average value of the square norm
of the statevector is not zero, we have to add an extra term to
equation (\ref{nm1}), as expected and as it happens also in the
case of white noise. Relation (\ref{ie2}) tells us
which kind of term must be added. The conclusion follows: with
reference to our procedure, the request that $P_{\makebox{\tiny
Cook}}[w(t)]$ correctly defines a probability distribution, i.e.
that the average value of the square norm of the statevector
$|\psi(t)\rangle$ is conserved, leads to the stochastic
Schr\"odinger equation:
\begin{equation}\label{sch2}
\frac{d|\psi(t)\rangle}{dt} =  \left[ -\frac{i}{\hbar} H_{0}
+ \sum_{i} A_{i} w_{i}(t) -
2\gamma \sum_{i,j}A_{i} \int_{0}^{t} ds\, D_{ij}(t,s)
\frac{\delta}{\delta w_{j}(s)} \right] |\psi(t)\rangle.
\end{equation}
This is the main result of this section. Note that an equation
like (\ref{sch2}) has been derived in references \cite{gi2, bud}
by following a different line of thought.

Some comments are appropriate:
\begin{itemize}
\item Equation (\ref{sch2}) no longer describes a Markovian
evolution for the sta\-te\-vector unless the correlation functions
$D_{ij}(t,s)$ are Dirac--$\delta$'s in the time variable ---
i.e. the stochastic processes $w_{i}(t)$ are white in time. As a
consequence, the corresponding equation for the statistical
operator is not of the quantum--dynamical--semigroup type
\cite{lin, bgrw}, contrary to what happen for the case of CSL
(see equation (\ref{cslso})).

\item In general,  the explicit form of the functional derivatives
of $|\psi(t)\rangle$ with respect to the noise $w_{i}(t)$ cannot
be evaluated exactly, except for few special cases, two of which
will be considered in the next section. Therefore, in the general
case it is difficult to analyze the time evolution of the
statevector and the statistical properties of the ensemble of
states generated by the stochastic processes. In particular, one
cannot write a closed equation for the evolution of the
statistical operator.
\end{itemize}

\section{Two special cases}

In order to understand the kind of difficulties one encounters
when working with non--white stochastic processes, and in
particular the reasons for which the functional derivative of the
statevector $|\psi(t)\rangle$ in general cannot be computed
exactly, let us reconsider equation (\ref{fs}), writing explicitly
its perturbative expansion:
\begin{eqnarray} \label{fs2}
\lefteqn{ T\, e^{\displaystyle -\, \frac{i}{\hbar}H_{0}t\, +
\sum_{i} A_{i} \int_{0}^{t} w_{i}(s)\, ds } \; =} \qquad\qquad
\nonumber \\
& = &
\sum_{n = 0}^{\infty} \left[-\frac{i}{\hbar}\right]^{n}\,
\frac{1}{n!} \,
\int_{0}^{t} dt_{1} \cdots \int_{0}^{t} dt_{n}\, T\left\{H(t_{1})
\ldots H(t_{n}) \right\}, \quad
\end{eqnarray}
where we have defined the operator:
\begin{equation} \label{yhj}
H(t) \quad = \quad H_{0} \; + \; i\hbar\sum_{i}A_{i}\,w_{i}(t).
\end{equation}

The functional derivative of $|\psi(t)\rangle$ with respect to
$w_{j}(s)$ can be obtained deriving  term by term the
series\footnote{We assume that the initial state $|\psi(0)\rangle$
does not depend on the stochastic processes $w_{i}(t)$.}
(\ref{fs2}). The derivative of the term $n=0$ is zero; the
derivative of the term $n=1$ is:
\begin{equation} \label{topd}
\frac{\delta}{\delta w_{j}(s)} \left[ -\frac{i}{\hbar}\,
\int_{0}^{t} dt_{1}\, H(t_{1}) \right] \; = \;
-\frac{i}{\hbar}\, \int_{0}^{t} dt_{1}\, \left[ i\hbar\, \delta(s
- t_{1})\, A_{j} \right] \; = \; A_{j}.
\end{equation}
The next ($n=2$) term is:
\begin{equation}
\left[-\frac{i}{\hbar}\right]^{2}\, \frac{1}{2}\,
\int_{0}^{t} dt_{1} \int_{0}^{t} dt_{2}\, T\left\{H(t_{1})
\, H(t_{2}) \right\}.
\end{equation}
The functional derivative of the time--ordered
product $T \{ H(t_{1})\, H(t_{2}) \} =
\theta(t_{1}-t_{2})H(t_{1})H(t_{2}) + \theta(t_{2} - t_{1})
H(t_{2}) H(t_{1})$ is:
\begin{eqnarray} \label{dftt}
\lefteqn{\frac{\delta}{\delta w_{j}(s)} T\left\{ H(t_{1}) \,
H(t_{2}) \right\} \; = \qquad\quad} \nonumber \\
& = &
i\hbar\,\theta(t_{1}-t_{2})\left[ \delta(t_{1} - s)\, A_{j}\,
H(t_{2}) \; + \; \delta(t_{2} - s)\, H(t_{1})\, A_{j}
\right] \nonumber \\
& + &
i\hbar\,\theta(t_{2}-t_{1})\left[ \delta(t_{2} - s)\, A_{j}\,
H(t_{1}) \; + \; \delta(t_{1} - s)\, H(t_{2})\, A_{j}
\right].
\end{eqnarray}
We note that the first and third terms at the right hand side of
(\ref{dftt}) differ only for the exchange of the dummy variables
$t_{1} \leftrightarrow t_{2}$; the same is true for the second and
the fourth term. The derivative of the $n=2$ term (i.e. of Eq. (4.4))
is
then:
\begin{equation} \label{rftt}
A_{j} \left[ -\frac{i}{\hbar}\, \int_{0}^{s} dt_{1}\,
H(t_{1}) \right] \; + \;
\left[ -\frac{i}{\hbar}\, \int_{s}^{t} dt_{1}\,
H(t_{1}) \right] A_{j}.
\end{equation}
Equation (\ref{rftt}) does not have a simple form, contrary to
(\ref{topd}), and derivatives of higher terms are more and more
complicated, due to the fact that the operators $A_{j}$ in general
do not commute with the Hamiltonian $H_{0}$. In fact, would they
commute, equation (\ref{rftt}) would simplify to:
\begin{equation}
A_{j} \left[ -\frac{i}{\hbar}\, \int_{0}^{t} dt_{1}\,
H(t_{1}) \right],
\end{equation}
i.e. the derivative of the second term would give $A_{j}$ times
the first term. Moreover, if $[A_{j}, H_{0}] = 0$, the
functional derivative of the term $n+1$ gives $A_{j}$ times the
$n$--th term:
\begin{equation} \label{nhlqr}
\frac{\delta}{\delta w_{j}(s)}\, |\psi(t)\rangle \quad = \quad
A_{j}\, |\psi(t)\rangle,
\end{equation}
as we are going to prove. In fact, the hypothesis
that the operators $A_{i}$ commute with the Hamiltonian $H_{0}$ is
equivalent to the (more elegant) requirement that the operators
$H(t)$ defined in (\ref{yhj}) commute at different times. In this
case, the time--ordered product in the exponential series
(\ref{fs2}) can be omitted, and the functional derivative of the
$n$--th term is:
\begin{eqnarray}
\lefteqn{\frac{\delta}{\delta\, w_{j}(s)} \left[-\frac{i}{\hbar}
\right]^{n}\, \frac{1}{n!} \, \int_{0}^{t} dt_{1} \cdots
\int_{0}^{t} dt_{n}\, \left\{H(t_{1}) \ldots H(t_{n})
\right\} \; =}
\nonumber \\
& = &
\left[-\frac{i}{\hbar}\right]^{n}\, \frac{1}{n!} \sum_{i=1}^{n}
\int_{0}^{t} dt_{1} \cdots  \int_{0}^{t} dt_{n}\, \left\{H(t_{1})
\ldots \frac{\delta\, H(t_{i})}{\delta\ w_{j}(s)} \ldots H(t_{n})
\right\}\; = \nonumber \\
& = &
\left[-\frac{i}{\hbar}\right]^{n}\, \frac{1}{(n-1)!}
\int_{0}^{t} dt_{1} \cdots  \int_{0}^{t} dt_{n}\, \left\{
\frac{\delta\, H(t_{1})}{\delta\ w_{j}(s)} \ldots H(t_{n})
\right\}\; = \nonumber \\
& = &
A_{j} \left[-\frac{i}{\hbar}\right]^{n-1}\, \frac{1}{(n-1)!}
\int_{0}^{t} dt_{1} \cdots  \int_{0}^{t} dt_{n-1}\, \left\{
H(t_{1}) \ldots H(t_{n-1})
\right\}.
\end{eqnarray}
This completes the proof. Note also that, when $s=t$, an extra
factor $1/2$ appears in (\ref{nhlqr}), because in this case the
Dirac delta function arising from the functional derivative of
$H(t)$ is centered in one of the two extreme points of the
interval of integration.

Recently, S. Adler and P. Horwitz \cite{ad1} (see also \cite{ad2})
have proposed a white--noise model of dynamical reductions in
which the operators $A_{i}$ are taken to be functions of the
Hamiltonian $H_{0}$; this implies that the stochastic terms of
equation (\ref{csl1}) drive the statevector into the {\it energy}
eigenmanifolds of the physical system.
Making such a choice in the non--white equation (\ref {sch2}), the
operators $H(t)$ at different times commute among themselves, the
functional derivatives of the statevector $|\psi(t)\rangle$ can be
computed, and equation (\ref{sch2}) becomes:
\begin{equation}\label{sch2a}
\frac{d|\psi(t)\rangle}{dt} =  \left[ -\frac{i}{\hbar} H_{0}
+ \sum_{i} A_{i} w_{i}(t) -
2\gamma \sum_{i,j}A_{i} A_{j} \int_{0}^{t}
D_{ij}(t,s)\,ds \right] |\psi(t)\rangle,
\end{equation}
with $A_{i} = A_{i}(H_{0})$. Equation (\ref{sch2a}) is exact and,
correspondingly, one can easily derive a closed equation for the
time evolution of the statistical operator. All the statistical
properties concerning the physical system can be evaluated
exactly.

We conclude the section showing that the functional derivatives of
$|\psi(t)\rangle$ can be explicitly evaluated also in the case of
general white noise stochastic processes,  without having to require
that $H_{0}$ commutes with $A_{i}$. Moreover, we will prove that in
this case equation (\ref{sch2}) reduces to (\ref{csl1}), as
expected.

Under the assumption of white--noise stochastic processes
($D_{ij}(t_{1}, t_{2}) \quad = \quad \delta_{ij}\,\delta(t_{1} -
t_{2})$), the Furutsu--Novikov relation
\begin{equation} \label{fnwn}
\llangle F[w(t)]w_{i}(t) \rrangle \quad = \quad
\gamma
\LLangle \frac{\delta F[w(t)]}{\delta w_{i}(t)} \RRangle
\end{equation}
leads to the following expression for the time derivative of the
average value of the square norm of the statevector $|\psi(t)\rangle$
satisfying equation (\ref{nm1}):
\begin{eqnarray} \label{ie2wn}
\frac{d}{dt}\llangle\langle\psi(t)|\psi(t)\rangle\rrangle & = &
2\gamma \sum_{i}
\LLangle \left[ \frac{\delta \langle\psi(t)|}{\delta
w_{i}(t)}\right] A_{i}|\psi(t)\rangle\RRangle  \; +
\nonumber \\  &  &
2\gamma \sum_{i}
\LLangle \langle\psi(t)|A_{i} \left[ \frac{\delta
|\psi(t)\rangle}{\delta w_{i}(t)} \right]\RRangle.
\end{eqnarray}
We now have to evaluate the functional derivatives of the
statevector, taking into account that the noises $w_{i}(t)$
(appearing in the derivatives) are taken at time $t$.

The derivative of the term $n=1$ is equal to $(1/2)A_{j}$ (see
equation (\ref{topd})), the factor $(1/2)$ deriving from the Dirac
delta
function
$\delta(t - t_{1})$ which is integrated between $0$ and $t$. For
the derivative of the $n=2$ term, let us look at expression
(\ref{rftt}). If we take $s=t$, the second term goes to zero,
while the first one gives\footnote{The factor $(1/2)$ appears for
the same reason as before.}:
\begin{equation}
\frac{1}{2}\, A_{j} \left[ -\frac{i}{\hbar} \int_{0}^{t} dt_{1}\,
H(t_{1}) \right].
\end{equation}
In general, the functional derivative of any terms of the
exponential series (\ref{fs2}) gives $(1/2)A_{j}$ times the
previous term, so that:
\begin{equation} \label{rsncwn}
\frac{\delta}{\delta w_{j}(t)}\, |\psi(t)\rangle \quad = \quad
\frac{1}{2}\,A_{j}\, |\psi(t)\rangle.
\end{equation}
This means that the square--norm--preserving Schr\"odinger
equation is:
\begin{equation}\label{sch3}
\frac{d|\psi(t)\rangle}{dt} =  \left[ -\frac{i}{\hbar} H_{0}
+ \sum_{i} A_{i}w_{i}(t) -
\gamma \sum_{i} \,A_{i}\, A_{j} \right] |\psi(t)\rangle,
\end{equation}
which coincides with the original CSL equation (\ref{csl1}). An
alternative and quicker way to derive the white--noise limit is
to replace $D_{ij}(t,s)$ with $\delta_{ij} \delta(t-s)$ in
equation (\ref{sch2}) and to show that (\ref{rsncwn}) is a
consistent solution.

\section{The reduction mechanism}

Here, we will analyze under which conditions the new terms in the
modified Schr\"odinger equation (\ref{sch2}) induce, for large
times, the reduction of the statevector to one of the common
eigenstates of the commuting operators $A_{i}$.

For this purpose, let us disregard the Hamiltonian $H_{0}$; under
this assumption the operators $H(t)$ commute at different times
and (as discussed in the previous section) the functional
derivatives of the statevector $|\psi(t)\rangle$ give the
operators $A_{i}$ times $|\psi(t)\rangle$. Equation (\ref{sch2})
becomes then\footnote{Here and in what follows, we consider a
generic initial time $t_{0}$.}:
\begin{equation}\label{schred}
\frac{d|\psi(t)\rangle}{dt} =  \left[ \sum_{i} A_{i}\, w_{i}(t) -
2\gamma \sum_{i,j}A_{i} A_{j} \int_{t_{0}}^{t}
D_{ij}(t,s)\,ds \right] |\psi(t)\rangle.
\end{equation}
The equation for the statistical operator can now be easily
derived; using the definition (\ref{rcdos}), we get:
\begin{equation} \label{erops}
\frac{d \rho(t)}{dt} \quad = \quad -\gamma \sum_{i,j}
\left[ A_{i}, \left[ A_{j}, \rho(t)\right]\right]
\int_{t_{0}}^{t} D_{ij}(t,s)\, ds,
\end{equation}
which is a consistent generalization of the CSL equation
(\ref{cslso}) when the Hamiltonian $H_{0}$ is omitted: in fact,
if the stochastic processes $w_{i}(t)$ are independent and
white ($D_{ij}(t_{1}, t_{2}) = \delta_{ij}\, \delta(t_{1} -
t_{2})$), then (\ref{erops}) reduces exactly to (\ref{cslso}).

In order to test the reduction properties, we will show first of
all how the reduction mechanism works for the statistical operator
(see footnote 1). As in section 2, let us suppose that the common
eigenmanifolds of the operators $A_{i}$, which we assume to have a
purely discrete spectrum, are one--dimensional; let
$|\alpha\rangle$ be the vector spanning the
$\alpha$--eigenmanifold. The equation for the matrix elements
$\langle\alpha|\rho(t)|\beta\rangle$ is:
\begin{equation} \label{dada}
\frac{d \langle\alpha|\rho(t)|\beta\rangle}{dt} \; = \;
- \, \gamma \sum_{i,j} (a_{i\alpha} - a_{i\beta})(a_{j\alpha} -
a_{j\beta}) \int_{t_{0}}^{t} D_{ij}(t,s)\, ds \,
\langle\alpha|\rho(t)|\beta\rangle.
\end{equation}
Making use of the symmetry property of the correlation functions:
\begin{equation}
D_{ij}(t_{1}, t_{2}) \quad = \quad
D_{ji}(t_{2}, t_{1}),
\end{equation}
we can write the solution of equation (\ref{dada}) in the
following form (see also \cite{rgt}):
\begin{equation} \label{sesr}
\langle\alpha|\rho(t)|\beta\rangle  =  e^{\displaystyle - \,
\frac{\gamma}{2} \sum_{i,j} (a_{i\alpha} - a_{i\beta})(a_{j\alpha}
- a_{j\beta}) \int_{t_{0}}^{t} dt_{1} \int_{t_{0}}^{t} dt_{2}\,
D_{ij}(t_{1},t_{2})} \langle\alpha|\rho(t_{0})|\beta\rangle.
\end{equation}
From equation (\ref{sesr}), we sees that if $|\alpha\rangle =
|\beta\rangle$, the exponent is zero: as in CSL, the diagonal
elements of the density matrix do not change in time. If, on other
other hand $|\alpha\rangle \neq |\beta\rangle$, the evolution of
the matrix element depends on the time behavior the correlation
functions $D_{ij}(t_{1},t_{2})$.

If we want the off--diagonal elements to be damped at large times,
two conditions must be satisfied. The first one is that the {\bf
exponent} in (\ref{sesr}) must be {\bf negative}: this is always
true, since the correlation function of a Gaussian process is
positive definite.

The second condition is that the {\bf double integral} of the
correlation function must {\bf diverge} for large times:
\begin{equation} \label{cond2}
\int_{t_{0}}^{t} dt_{1} \int_{t_{0}}^{t} dt_{2}\,
D_{ij}(t_{1},t_{2}) \; \longrightarrow \; +\infty \qquad
\makebox{for $t \rightarrow +\infty$},
\end{equation}
so that the off--diagonal elements of the density matrix go to
zero. This condition is not {\it a priori} satisfied by a generic
Gaussian stochastic field. At any rate, physical reasonable
stochastic fields always satisfy it: here we present just a couple
of meaningful examples.

Suppose the stochastic fields $w_{i}(t)$ are equal and
independent, with a (normalized) Gaussian correlation function:
\begin{equation} \label{gcf}
D_{ij}(t_{1},t_{2}) \quad = \quad \delta_{ij}\,
\frac{1}{\sqrt{2\pi}\tau}\,e^{\displaystyle - \frac{(t_{1} -
t_{2})^{2}}{2\tau^{2}}}.
\end{equation}
Let us also take $t_{0} = -\infty$. Equation (\ref{dada}) then
becomes:
\begin{equation} \label{dadae1}
\frac{d \langle\alpha|\rho(t)|\beta\rangle}{dt} \; = \;
- \, \frac{\gamma}{2} \sum_{i} (a_{i\alpha} - a_{i\beta})^{2} \,
\langle\alpha|\rho(t)|\beta\rangle,
\end{equation}
which is independent from the correlation time $\tau$, and
moreover it corresponds exactly to the CSL equation (\ref{rm}).
The fact that we have taken $t_{0} = -\infty$ means that the
correspondence between equation (\ref{dada}) --- with a
correlation function like (\ref{gcf}) --- and equation (\ref{rm})
is exact only in the limit of large times. Note also that if we
take the limit $\tau \rightarrow 0$, the gaussian process becomes
a white noise process with a Dirac--$\delta$ correlation function and
we
recover, again, the CSL theory.

As a second example, suppose the correlation function is:
\begin{equation} \label{gcf2}
D_{ij}(t_{1},t_{2}) \quad = \quad \delta_{ij}\,
\frac{1}{2\tau}\,e^{\displaystyle - \frac{|t_{1} -
t_{2}|}{\tau}}.
\end{equation}
Equation (\ref{dada}) becomes:
\begin{equation} \label{dadae2}
\frac{d \langle\alpha|\rho(t)|\beta\rangle}{dt} \; = \; - \,
\frac{\gamma}{2} \left[1- e^{\displaystyle -\frac{(t -
t_{0})}{\tau}}\right] \sum_{i} (a_{i\alpha} - a_{i\beta})^{2} \,
\langle\alpha|\rho(t)|\beta\rangle.
\end{equation}
As before, the off--diagonal elements are exponentially damped
and, in the limit $t \rightarrow +\infty$ we recover the behavior
of CSL. Note that the effect of a non--white correlation function
is that of decreasing the reduction rate of the localization
mechanism.

We now analyze how the reduction mechanism works at the
wavefunction level, proving in this way that equation
(\ref{schred}) leads to the reduction of the statevector into one
of the common eigenmanifolds of the operators $A_{i}$. As in
reference \cite{ggb}, we consider a simplified dynamics in which
only one operator $A$ appears in equation (\ref{schred}). This
operator is coupled to a single stochastic process $w(t)$, whose
correlation function is $D(t_{1}, t_{2})$. Finally, we assume that
at the initial time $t_{0}$ the statevector is:
\begin{equation}
|\psi(t_{0})\rangle \quad = \quad P_{\alpha} |\psi(t_{0})\rangle
\; + \; P_{\beta} |\psi(t_{0})\rangle,
\end{equation}
where $P_{\alpha}$ and $P_{\beta}$ are projection operators onto
the eigenmanifolds of $A$ corresponding to two different
eigenvalues $\alpha$ and $\beta$, respectively. The solution of
equation (\ref{schred}) is:
\begin{equation}
|\psi(t)\rangle \; = \; e^{\displaystyle \alpha x(t) - \alpha^{2}
\gamma f(t)}P_{\alpha} |\psi(t_{0})\rangle \; + \; e^{\displaystyle
\beta x(t) - \beta^{2} \gamma f(t)}P_{\alpha} |\psi(t_{0})\rangle,
\end{equation}
where
\begin{equation}
x(t) \; = \; \int_{t_{0}}^{t} w(s)\, ds, \qquad\quad
f(t) \; = \; \int_{t_{0}}^{t} ds_{1} \int_{t_{0}}^{t}
ds_{2}\, D(s_{1}, s_{2}).
\end{equation}
Note that $\gamma f(t) = \llangle x^{2}(t) \rrangle$, i.e. such a
quantity is the variance of the stochastic process $x(t)$.

Since the ``raw'' probability distribution of the process $x(t)$
is:
\begin{equation}
P_{\makebox{\tiny Raw}}[x(t)] \quad  = \quad \frac{1}{\sqrt{2\pi
\gamma f(t)}}\, e^{\displaystyle -\frac{1}{2\gamma f(t)}\,
x^{2}(t)},
\end{equation}
taking into account the cooking prescription (\ref{prob2}) we
obtain:
\begin{eqnarray} \label{nmvd}
P_{\makebox{\tiny Cook}}[x(t)] &  = & \|P_{\alpha}
|\psi(t_{0})\rangle \|^{2}\, \frac{1}{\sqrt{2\pi\gamma f(t)}}\,
e^{\displaystyle -\frac{1}{2\gamma f(t)}\, [x(t) - 2\alpha\gamma
f(t)]^{2}} \nonumber \\
&  + & \|P_{\beta} |\psi(t_{0})\rangle \|^{2}\,
\frac{1}{\sqrt{2\pi\gamma f(t)}}\, e^{\displaystyle -\frac{1}{2\gamma
f(t)}\, [x(t) - 2\beta\gamma f(t)]^{2}}. \nonumber \\
\end{eqnarray}
Equation (\ref{nmvd}) implies that, if $f(t) \rightarrow +\infty$
when $t \rightarrow +\infty$, the stochastic process $x(t)$ will
take either a value close to $2\alpha\gamma f(t)$ --- within an
interval of width $\sqrt{\gamma f(t)}$ --- or a value close to
$2\beta\gamma f(t)$, within the same interval\footnote{As noted in
\cite{ggb}, even though the interval $\sqrt{\gamma f(t)}$ tends to
infinity as time increases, the ratio $\sqrt{\gamma f(t)}/
2(\alpha - \beta)\gamma f(t)$ goes to zero.}. Of course, the
requirement that $f(t) \rightarrow +\infty$ as time increases is
exactly the same as requirement (\ref{cond2}) which guarantees the
damping of the off--diagonal elements of the density matrix.

Suppose now that the actual realization of the stochastic process
$x(t)$ occurs around $2\alpha\gamma f(t)$; the corresponding
probability is $\|P_{\alpha} |\psi(0)\rangle \|^{2}$. We then
have:
\begin{equation}
\frac{\|P_{\beta} |\psi(t)\rangle \|^{2}}{\|P_{\alpha}
|\psi(t)\rangle \|^{2}} \; \simeq \; e^{\displaystyle
-2\gamma(\alpha - \beta)^{2} f(t)} \frac{\|P_{\beta}
|\psi(0)\rangle \|^{2}}{\|P_{\alpha} |\psi(0)\rangle \|^{2}} \;
\rightarrow \; 0 \qquad \makebox{as $t \rightarrow \infty$},
\end{equation}
which means that the statevector $|\psi(t)\rangle$ is driven into
the eigenmanifold of the operator $A$ corresponding to the
eigenvalue $\alpha$. By the same reasoning, it is immediate to see
that, with a probability equal to $\|P_{\beta} |\psi(0)\rangle
\|^{2}$, the statevector is driven into the eigenmanifold
associated to the eigenvalue $\beta$. We have thus proved that the
statevector $|\psi(t)\rangle$  undergoes a random spontaneous
localization into one of the two eigenmanifolds of the operator
$A$, with a probability which coincides with the one assigned by
standard Quantum Mechanics to the outcomes of an experiment aimed
to measure the observable $A$.

\section{The average value of observables}

When one disregards the Hamiltonian term $H_{0}$, it is not
difficult to see how the stochastic terms affect the average
value of physical quantities.

The mean value of an operator $O$ (for simplicity, we consider an
observable which does not depend explicitly on time) is defined as
the expectation value $\langle\phi(t)|O|\phi(t)\rangle$, averaged
over all possible realizations of the stochastic noises:
\begin{eqnarray*}
\langle O\rangle  & = &
\int {\mathcal D}[w(t)]\, \langle\psi(t)| O |\psi(t)\rangle
\, P_{\makebox{\tiny Cook}}[w(t)] \quad = \\
& = & \llangle \langle\psi(t)| O |\psi(t)\rangle \rrangle.
\end{eqnarray*}
Its time derivative can be calculated following almost the same
steps which, in the previous section, have led to equation
(\ref{erops}) for statistical operator; the final equation is:
\begin{equation} \label{av}
\frac{d\langle O\rangle}{dt} \quad  = \quad -\gamma
\sum_{i,j} \llangle \langle\psi(t)|
\left[ A_{i},\left[ A_{j}, O \right]\right]
|\psi(t)\rangle \rrangle
\int_{t_{0}}^{t} D_{ij}(t,s)\, ds,
\end{equation}
to be compared with the corresponding CSL--white noise equation:
\begin{equation} \label{avcsl}
\frac{d\langle O\rangle}{dt} \quad  = \quad -\,\frac{\gamma}{2}
\sum_{i} \llangle \langle\psi(t)|
\left[ A_{i},\left[ A_{i}, O \right]\right]
|\psi(t)\rangle \rrangle.
\end{equation}
The analysis of the previous section should have made clear how
(\ref{av}) differs from (\ref{avcsl}), so we will not repeat it
here.

\section{Connection with CSL}

We now apply the formalism introduced in the previous sections to
derive an equation with the property of localizing macroscopic
systems in {\it space}, like in CSL; in other words, we specify
the choice of the ``preferred basis'' $\{ A_{i} \}$ in such a way to
have
a physically meaningful theory for our purposes.

The most natural choice for the operators $A_{i}$ is the number
density operator for a system of identical particles:
\begin{equation}
A_{i} \quad \longrightarrow \quad N({\bf x}) \; = \; \sum_{s}
a^{\dagger}({\bf x}, s)\,a({\bf x}, s).
\end{equation}
Correspondingly, the noises $w_{i}(t)$ are replaced by a
stochastic field $w({\bf x}, t)$, whose correlation function is
$D({\bf x}, t_{1}; {\bf y}, t_{2})$.

In reference \cite{ggp}, the transformation and invariance
properties of dynamical reduction models have been discussed in
detail. In particular, it has been proved that, in order for the
physical properties of the model to be invariant under Galilean
transformations (we speak of {\it stochastic Galilean
invariance}), the correlation function $D({\bf x}, t_{1}; {\bf y},
t_{2})$ itself must be invariant under the considered group of
transformations, i.e.
\begin{equation} \label{fhf}
D({\bf x}, t_{1}; {\bf y}, t_{2}) \quad = \quad
D(|{\bf x} - {\bf y}|, t_{1}- t_{2});
\end{equation}
the easiest way to construct a function like (\ref{fhf}) is to
take the product of two functions of the space and time
variables, respectively:
\begin{equation}
D({\bf x}, t_{1}; {\bf y}, t_{2}) \quad = \quad g(|{\bf x} - {\bf
y}|)\, h(t_{1} - t_{2}).
\end{equation}
As regards $g(|{\bf x} - {\bf y}|)$, a reasonable choice is a
gaussian function, like in CSL:
\begin{equation}
g(|{\bf x} - {\bf y}|) \quad = \quad \gamma
\left(\frac{\alpha}{4\pi}\right)^{\frac{3}{2}}\,
e^{\displaystyle - \frac{\alpha}{4} ({\bf x} - {\bf y})^{2}},
\end{equation}
with $1/\sqrt{\alpha} \simeq 10^{-5}$ cm.

It is  natural to choose a gaussian function also for
$h(t_{1} - t_{2})$:
\begin{equation}
h(t_{1} - t_{2}) \quad = \quad
\left(\frac{\beta}{4\pi}\right)^{\frac{1}{2}}\,
e^{\displaystyle - \frac{\beta}{4} (t_{1} - t_{2})^{2}}.
\end{equation}
With the above choice, we have introduced a new parameter
($\beta$); this can be considered as a drawback of the model.
However, we note that it always is possible to define $\beta$ in
terms of $\alpha$, $\gamma$ and fundamental constants of nature,
so that no new arbitrary parameter is introduced into the model.
As an example, we can choose $\beta = c^{2}\alpha \simeq 10^{30}$
sec${}^{-2}$, where $c$ is the speed of light. This choice is
particularly appropriate in the light of a possible relativistic
generalization of the theory, which we will discuss in a future
paper. Moreover, such a choice corresponds to an extremely small
correlation time, so that for ordinary systems (moving slower than
the speed of light) the behavior of the model is similar to the
one deriving from the white--noise CSL.

The modified equation (\ref{sch2}) for the statevector evolution
becomes now:
\begin{eqnarray} \label{ecnmr}
\frac{d |\psi(t)\rangle}{dt} & = & \left[ -\frac{i}{\hbar} H_{0}
\; + \; \int d^{3}x\, N{(\bf x}) w({\bf x}, t) \; - \right.
\\
& & \left. - \; 2\gamma \int d^{3}x d^{3}y\, N({\bf x})
g(|{\bf x} - {\bf y}|) \int_{t_{0}}^{t} ds\, h(t - s)
\frac{\delta}{\delta w({\bf y}, s)} \right]|\psi(t)\rangle.
\nonumber
\end{eqnarray}
If we ignore the free Hamiltonian $H_{0}$, i.e. we confine our
considerations to the reduction mechanism\footnote{For the physically
interesting cases, e.g. for the dynamical evolution of macrosystems,
such
an assumption is justified by the fact that the effect of the
reduction
is much faster that the tipical times in which the Hamiltonian can
induce
appreciable dynamical changes of the statevector.}, equation
(\ref{ecnmr})
becomes:
\begin{equation} \label{enmrsh}
\frac{d |\psi(t)\rangle}{dt} \; = \; \left[
\int d^{3}x\, N{(\bf x}) w({\bf x}, t) \; - \; \gamma(t)
\int d^{3}x d^{3}y\, N({\bf x})g(|{\bf x} - {\bf y}|)
N({\bf y}) \right] |\psi(t)\rangle
\end{equation}
with:
\begin{equation}
\gamma(t) \quad = \quad 2\gamma \int_{t_{0}}^{t} ds\,
h(t - s).
\end{equation}
The corresponding equation for the statistical operator is:
\begin{equation} \label{eosnsh}
\frac{d}{dt}\, \rho(t) \; = \; -\frac{\gamma(t)}{2}
\int d^{3}x\, d^{3}y
\left[ N({\bf x}), \left[ N({\bf y}), \rho(t) \right] \right]
g(|{\bf x} - {\bf y}|).
\end{equation}

Equation (\ref{ecnmr}) can be rewritten in a form closer to
equation (\ref{sch1b}), which will be useful for the subsequent
discussion. Let us define a new Gaussian stochastic process
$\overline{w}({\bf x}, t)$, which is connected to $w({\bf x}, t)$
by the relation:
\begin{equation}
w({\bf x}, t) \quad = \quad
\left(\frac{\alpha}{2\pi}\right)^{\frac{3}{2}} \int d^{3}x\,
e^{\displaystyle -\frac{\alpha}{2} ({\bf x} - {\bf y})^{2}}
\overline{w}({\bf y}, t).
\end{equation}
$\overline{w}({\bf x}, t)$ has zero mean, and correlation function
\begin{equation}
\llangle \overline{w}({\bf x}, t_{1})\,
\overline{w}({\bf y}, t_{2}) \rrangle \quad = \quad
\gamma\, \delta^{(3)}({\bf x} - {\bf y})\,
h(t_{1} - t_{2}).
\end{equation}
Using the following relation:
\begin{eqnarray}
\frac{\delta}{\delta \overline{w}({\bf x}, s)}\, |\psi(t)\rangle
& = &
\int d^{3}y\; \frac{\delta w({\bf y}, s)}{\delta \overline{w}({\bf x},
s)}\, \frac{\delta}{\delta w({\bf y}, s)}\, |\psi(t)\rangle
\quad = \nonumber \\
& = &
\left(\frac{\alpha}{2\pi}\right)^{\frac{3}{2}} \int d^{3}y\;
e^{\displaystyle -\frac{\alpha}{2} ({\bf x} - {\bf y})^{2}}
\frac{\delta}{\delta w({\bf y}, s)}\, |\psi(t)\rangle,
\end{eqnarray}
it can be easily seen that (\ref{ecnmr}) is equivalent to the
 equation:
\begin{eqnarray} \label{ecnmr2}
\frac{d |\psi(t)\rangle}{dt} & = & \left[ -\frac{i}{\hbar} H_{0}
\; + \; \int d^{3}x\, {\mathcal N}{(\bf x}) w({\bf x}, t) \; -
\right.
\\ & &
\left. - \; 2\gamma \int d^{3}x\, {\mathcal N}({\bf
x}) \int_{t_{0}}^{t} ds\, h(t - s) \frac{\delta}{\delta
\overline{w}({\bf x}, s)} \right]|\psi(t)\rangle, \nonumber
\end{eqnarray}
with ${\mathcal N}{(\bf x})$ defined by (\ref{cho2}).

\subsection{Dynamics for macroscopic rigid bodies}

As for white--noise CSL \cite{gpr}, it is not difficult to discuss
the physical implications of equation (\ref{ecnmr}) --- or
equation (\ref{ecnmr2}) --- for the case of a macroscopic rigid
body, i.e. a body such that the wavefunctions of its constituents
can be considered very well localized with respect to the
localization length $1/\sqrt{\alpha}$.

To be precise, in analogy with the procedure followed in
\cite{gpr}, let us consider a system of $N$ identical particles of
coordinates ${\bf q}_{i}$; let
\begin{equation}
{\bf Q} \quad = \quad \frac{1}{N}\, \sum_{i=1}^{N} {\bf q}_{i}
\end{equation}
be the center of mass coordinate, and let us write
\begin{equation}
{\bf q}_{i} \quad = \quad {\bf Q} \; + \; \overline{\bf q}_{i},
\end{equation}
where the coordinates $\overline{\bf q}_{i}$ are functions of $3N
- 3$ independent internal variables\footnote{See ref. \cite{gpr}
for further details on the degrees of freedom of the system.},
which we call $r$. Let us consider the wavefunction
\begin{equation}
|\psi(q,s)\rangle \; = \; |\phi({\bf Q})\rangle |\varphi
(r,s)\rangle \qquad\quad |\varphi (r,s)\rangle \; = \;
\left[
\begin{array}{c}
A \\ S
\end{array}
\right]
|\Delta(r,s)\rangle,
\end{equation}
where $q = \{ {\bf q}_{i} \}$ and $s = \{ s_{i} \}$ are the sets
of the space and spin coordinates of the $N$ particles,
respectively, while ``A'' and ``S'' mean symmetrization or
antisymmetrization with respect to the interchange of the
variables $({\bf q}_{i}, s_{i})$, respectively.

In reference \cite{gpr} it has been proved that if the
wavefunction of the internal degrees of freedom is very well
peaked with respect to the characteristic length
$1/\sqrt{\alpha}$, then, to an extremely high degree of accuracy,
\begin{equation}
{\mathcal N}({\bf x}) |\phi({\bf Q})\rangle |\varphi
(r,s)\rangle \quad = \quad F({\bf Q} - {\bf x})
|\phi({\bf Q})\rangle |\varphi (r,s)\rangle,
\end{equation}
with
\begin{equation} \label{dd}
F({\bf Q} - {\bf x}) \quad = \quad
\left(\frac{\alpha}{2\pi}\right)^{\frac{3}{2}} \sum_{i=1}^{N}
e^{\displaystyle -\frac{\alpha}{2} [{\bf Q} + \overline{\bf
q}_{i}(r_{0}) - {\bf x}]^{2}},
\end{equation}
where $r_{0}$ describes the set of the average equilibrium positions
of
the particles of the rigid body. Equation (\ref{dd}) means that the
operators ${\mathcal N}({\bf x})$ act only on the center of mass
wavefunction $|\phi({\bf Q})\rangle$.

As a consequence, if the Hamiltonian $H_{0}$ can be written as
\begin{equation}
H_{0} \quad = \quad H_{\bf Q} \; + \; H_{r},
\end{equation}
and if $|\phi({\bf Q})\rangle$ and $|\varphi (r,s)\rangle$ satisfy
the equations
\begin{eqnarray} \label{ecnmr3}
\frac{d |\phi({\bf Q}, t)\rangle}{dt} & = & \left[ -\frac{i}{\hbar}
H_{\bf Q} \; + \; \int d^{3}x\, {\mathcal N}{(\bf x})
\overline{w}({\bf x}, t) \; - \right.
\\ & &
\left. - \; 2\gamma \int d^{3}x\, {\mathcal N}({\bf
x}) \int_{t_{0}}^{t} ds\, h(t - s) \frac{\delta}{\delta
\overline{w}({\bf x}, s)} \right]|\phi({\bf Q}, t)\rangle, \nonumber
\\
& & \nonumber \\ \label{tyj}
\frac{d |\varphi (r,s,t)\rangle}{dt} & = & \left[ -\frac{i}{\hbar}
H_{r} \right] |\varphi (r,s,t)\rangle,
\end{eqnarray}
then $|\psi(q,s,t)\rangle$ satisfies equation (\ref{ecnmr2}) or,
equivalently, equation (\ref{ecnmr}).

Equations (\ref{ecnmr3}) and (\ref{tyj}) imply that the center of
mass and internal motion decouple, and that the stochastic terms
affect only the center of mass and not the internal structure, as
it happens for CSL.

Following the same arguments of reference \cite{gpr}, it can also
be proven that the localization rate of the center of mass
wavefunction grows linearly with the number of particles of the
rigid body. Such a localization rate can be easily computed  by
studying the time evolution of the off--diagonal elements $\langle
{\bf Q}'|\rho_{\bf Q}|{\bf Q}''\rangle$ of the reduced statistical
operator describing the center of mass motion of the system. As we
did in section 5, we disregard the Hamiltonian $H_{\bf Q}$, in
accordance
with  the fact that, for $|{\bf Q}' - {\bf Q}''| > 1/\sqrt{\alpha}$,
the
reduction rate  turns out to be much faster than the typical times
required in order that the standard Schr\"odinger evolution induces
appreciable changes of $|\phi({\bf Q}, t)\rangle$. Under this
assumption,
equation (\ref{ecnmr3}) becomes:
\begin{equation} \label{ppoi}
\frac{d |\phi({\bf Q}, t)\rangle}{dt} \; = \; \left[ \int d^{3}x\,
F({\bf Q} - {\bf x}) \overline{w}({\bf x}, t) - \gamma(t) \int
d^{3}x\, F^{2}({\bf Q} - {\bf x}) \right]|\phi({\bf Q}, t)\rangle,
\end{equation}
and the corresponding equation for the matrix elements $\langle {\bf
Q}'|\rho_{\bf Q}|{\bf Q}''\rangle$ is:
\begin{equation}
\frac{d \langle {\bf Q}'|\rho_{\bf Q}(t)|{\bf Q}''\rangle}{dt} \quad
= \quad - \Gamma({\bf Q}', {\bf Q}'', t) \,
\langle {\bf Q}'|\rho_{\bf Q}(t)|{\bf Q}''\rangle
\end{equation}
with
\begin{eqnarray}
\Gamma({\bf Q}', {\bf Q}'', t) & = & \gamma(t) \int d^{3}x\,
\left[ \frac{1}{2} F^{2}({\bf Q}' - {\bf x}) + \frac{1}{2}
F^{2}({\bf Q}'' - {\bf x}) - \right. \nonumber \\
& & \left. \frac{}{} F({\bf Q}' - {\bf x})
F({\bf Q}'' - {\bf x}) \right].
\end{eqnarray}
This is the same term as the one appearing in CSL, with
$\gamma(t)$ replacing $\gamma$; this proves that also in our model
the reduction frequency of the center of mass of the system grows
linearly with the number of its constituents. Moreover, taking a
large value for $\beta$, as it has been suggested previously,
$\gamma(t) \rightarrow \gamma$ in very short times, so that the
reducing dynamics is practically the same as the one of CSL.

\section*{Acknowledgements}

The authors would like to thank S. Adler for useful discussion and
advices.

\end{document}